\documentclass{article}                     
\usepackage{graphicx}
\usepackage{dcolumn}
\usepackage{bm}
\usepackage{amsmath}
\usepackage{graphicx}
\usepackage{amsmath,amssymb}

\newcommand{\mean}[1]{\langle{#1}\rangle}

\newcommand{\Tr}{{\rm Tr}\hspace{0.07cm}}

\newcommand{\half}{\frac{1}{2}}

\begin{document}

\title{ {\bf Entanglement-assisted quantum feedback control}
\thanks{
This work was supported in part by JSPS Grant-in-Aid No. 15K06151 
and JST PRESTO No. JPMJPR166A. 
The authors acknowledge helpful discussions with M. R. Hush, 
A. R. R. Carvalho, and S. S. Szigeti. }
}


\author{Naoki Yamamoto ~ and ~ Tomoaki Mikami 
\\
\\
Department of Applied Physics and Physico-Informatics, Keio University,  \\
              Hiyoshi 3-14-1, Kohoku, Yokohama 223-8522, Japan
}




\maketitle

\begin{abstract}
The main advantage of quantum metrology relies on the effective use of 
entanglement, which indeed allows us to achieve strictly better estimation 
performance over the standard quantum limit. 
In this paper, we propose an analogous method utilizing entanglement for the 
purpose of feedback control. 
The system considered is a general linear dynamical quantum system, where 
the control goal can be systematically formulated as a linear quadratic Gaussian 
control problem based on the quantum Kalman filtering method; 
in this setting, an entangled input probe field is effectively used to reduce 
the estimation error and accordingly the control cost function. 
In particular, we show that, in the problem of cooling an opto-mechanical 
oscillator, the entanglement-assisted feedback control can lower the stationary 
occupation number of the oscillator below the limit attainable by the controller 
with a coherent probe field and furthermore beats the controller with an optimized 
squeezed probe field. 

\end{abstract}


\section{Introduction}

\begin{figure}
\centering
\includegraphics[scale=0.34]{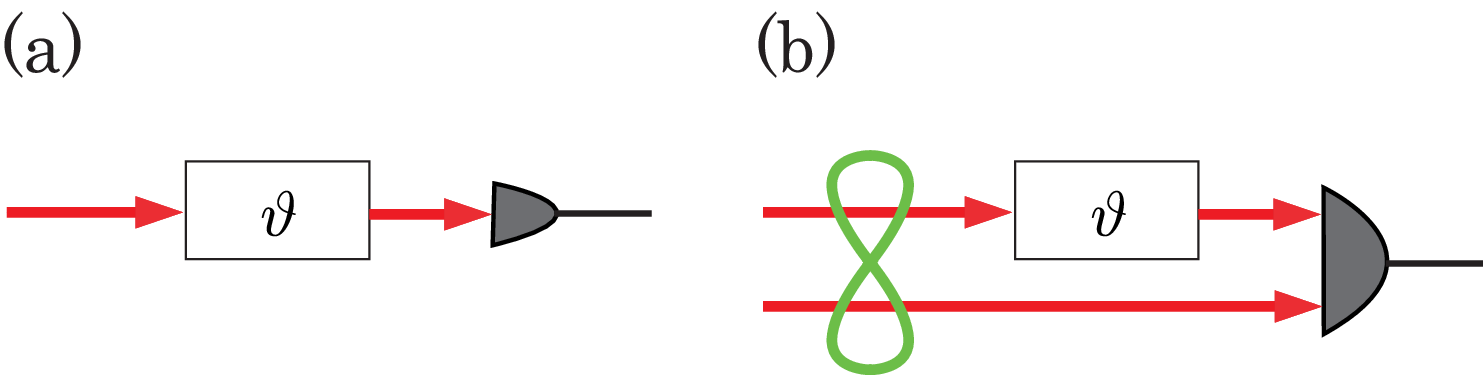}
\caption{\label{Q metrology}
(a) Standard setup for estimating an unknown parameter $\vartheta$. 
(b) Quantum metrology setup using an entangled input. 
}
\end{figure}

Entanglement is a special notion that had been considered as a ``spooky" 
correlation \cite{Einstein}. 
However over recent decades it has gained a positive impression mainly thanks 
to its central role in quantum information science \cite{Nielsen,Dowling&Milburn}. 
A particularly important application of entanglement in our context is the 
{\it quantum metrology} \cite{Braunstein,Giovanetti}. 
The basic configuration is depicted in Fig. \ref{Q metrology}. 
The goal is to estimate an unknown parameter $\vartheta$ of the system. 
A standard estimation method is first to send a known input state and 
then measure the output state containing the information about $\vartheta$ 
(Fig.~\ref{Q metrology}~(a)); 
in this case the estimation error has a strict lower bound called the 
{\it standard quantum limit (SQL)} with respect to the input energy. 
In the quantum metrology schematic depicted in Fig.~\ref{Q metrology}~(b), 
on the other hand, an entangled state is chosen as an input so that one 
portion passes through the system while the other portion does not; 
then by measuring the combined output, we obtain more information 
about $\vartheta$ than the standard case (Fig.~\ref{Q metrology} (a)) 
and thus can beat the SQL in the estimation error. 
This schematic have been experimentally demonstrated in several settings, 
e.g., \cite{Steinberg,Takeuchi,Polzik 2009,Polzik 2010}. 
Note that the entanglement-assisted method is not a unique approach for beating 
the SQL; particularly in the case where $\vartheta$ is the parameter of a force 
applied to a mechanical oscillator (e.g., a gravitational wave force), several 
alternative estimation schemes beating the SQL have been developed, such as 
the squeezed-probe scheme 
\cite{Kimble 2001,Aasi 2013,Furusawa 2013,Andersen 2016} and 
the variational measurement technique \cite{Vyatchanin 1995} for back-action 
evasion.

What we learn from the theory of quantum metrology is the fact that, in 
a broad sense, a quantum estimator could have better performance if 
assisted by entanglement. 
Therefore it is a reasonable idea to employ an entanglement-assisted 
estimation strategy for the measurement-based quantum feedback 
control \cite{WisemanBook,KurtBook}, which is now well established 
based on the quantum filtering theory \cite{Belavkin1999,Bouten}. 
Actually, in a similar configuration depicted in Fig.~\ref{Q metrology}~(b), 
it is expected that the quantum filter (i.e., the best continuous-time estimator) 
brings us more information, and as consequence we will have chance to 
construct a better controller than in the standard case without entanglement. 
The idea of entanglement-assisted feedback control is briefly mentioned 
in \cite{Genoni}, but there has been no quantitative analysis of this control 
strategy. 
That is, we are interested in the following questions; 
(i) {\it How much does the entanglement-assisted strategy improve the control 
performance in a realistic setup?} 
(ii) {\it In what situation is the entanglement-assisted feedback control really 
beneficial?} 
Note that the answers to these questions are non-trivial, because, for a realistic 
noisy system, the entanglement-assisted feedback control will not always 
outperform the standard one without entanglement, due to the fragile nature 
of entangled states.

\begin{figure}
\centering
\includegraphics[scale=0.36]{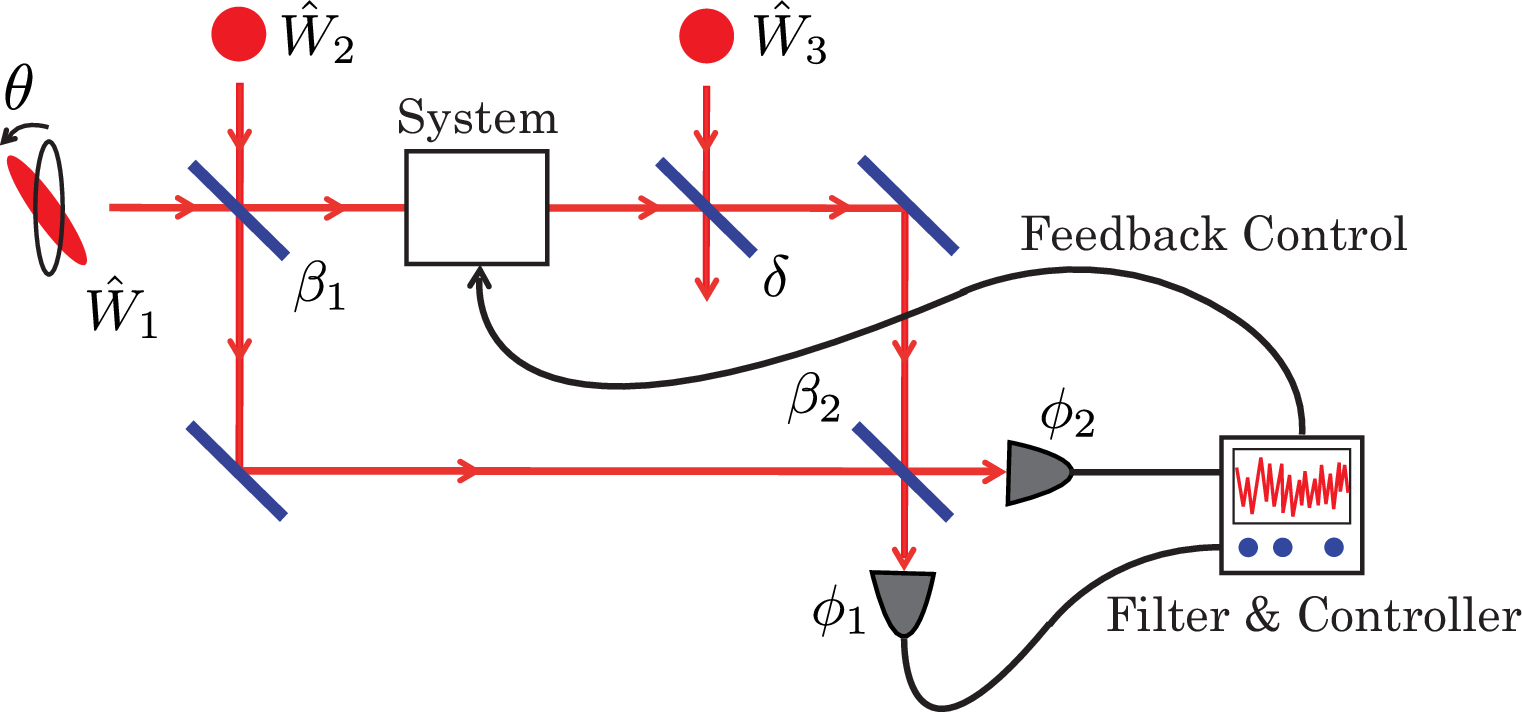}
\caption{\label{General setup}
General feedback control configuration via an entangled input field. 
$\hat W_1, \hat W_2$ are field annihilation operators representing 
a squeezed state with phase $\theta$ and a coherent (or vacuum) state, 
respectively. 
$\hat W_3$ represents a vacuum field. 
$\beta_1^2$, $\beta_2^2$, and $\delta^2$ denote the reflectivity of the beam 
splitters; in particular $\delta$ corresponds to an optical loss of the system's 
output field. 
$\phi_1$ and $\phi_2$ are the phases of Homodyne detectors. 
}
\end{figure}

In this paper, we consider the setup illustrated in Fig.~2. 
The system to be controlled is a general linear quantum system such as an 
optical amplifier and an opto-mechanical oscillator 
\cite{BachorBook,James2008,Nurdin 2009,Hamerly 2012,Yamamoto2014,Yamamoto2016}. 
The probe input is given by an optical entangled state generated by combining 
a squeezed field and a coherent field \cite{Furusawa2011} at a beam splitter; 
one portion of this entangled field couples to the system while the other portion 
does not, as in the scheme shown in Fig.~1 (b), and then the combined output 
field is continuously measured by Homodyne detectors. 
Finally, based on the measurement signal, we construct the quantum filter and 
then apply a feedback control to the system. 
The point of this setting is that the system state is always Gaussian, and as a 
result the quantum filter is simplified to the {\it quantum Kalman filter} 
\cite{Doherty1999a}, which enables us to compute the exact real-time estimate 
of the system variables. 
Furthermore, in this paper, we consider the quantum {\it Linear Quadratic 
Gaussian (LQG)} optimal control problem, meaning that the control goal is 
to minimize a quadratic-type cost function. 
Fortunately, again thanks to the Gaussianity of the system state, this problem 
can be analytically solved by almost the same way as in the classical case 
\cite{Doherty1999b,Belavkin2008}. 
An important fact in this formulation is that a strict lower bound of the 
cost, which is ideally achievable by employing the so-called {\it cheap control} 
\cite{Sivan,Seron Book,Seron}, is represented by a function of only the 
estimation error. 
Therefore this ultimate limit of the LQG control cost has the meaning of 
SQL, when the input is given by a coherent or a vacuum field.

In the above described framework, first, this paper proves that, thanks to 
the entangled probe input, the filter certainly gains additional information 
which may improve the control performance. 
Next we study a feedback cooling problem of an opto-mechanical oscillator 
\cite{Mancini 1998,Hopkins,Hamerly,Hofer 2015,Kippenberg 2015,Aspelmeyer 2015} 
and provide answers to the above-posed questions by conducting detailed 
numerical simulations. 
In particular, it is shown that, by carefully choosing the system parameters 
($\theta, \beta_1, \beta_2, \phi_1, \phi_2$; see Fig.~2), the 
entanglement-assisted feedback control can lower the stationary occupation 
number of the oscillator below the SQL in the sense of cheap control mentioned 
above, and moreover, it outperforms the control with an optimized squeezed 
probe field \cite{Andersen 2016}.

Finally, we note that the scheme presented in this paper differs from that 
studied in \cite{Hofer 2015}, which considers the use of system's {\it internal} 
entanglement to enhance cooling for an opto-mechanical system.

{\it Notation:} 
$\Re$ and $\Im$ denote the real and imaginary parts, respectively. 
$I_n$: $n\times n$ identity matrix. 
$O_n$: $n\times n$ zero matrix. 
$0_{n\times m}$: $n\times m$ zero matrix.


\section{Quantum Kalman filtering, LQG control, and cheap control}

We here review the general theory of quantum Kalman filtering, LQG 
optimal control, and the cheap control.


\subsection{Quantum linear systems}

In this paper we consider a linear quantum system, whose general form is 
described as follows (see \cite{WisemanBook,KurtBook,James2008,Yamamoto2014} 
for more details). 
The system variables are collected in a vector of operators 
$\hat x :=[\hat q_1, \hat p_1, \ldots, \hat q_n, \hat p_n]^\top$, 
where $\hat q_i$ and $\hat p_i$ are position and momentum operators. 
They satisfy the canonical commutation relation 
$\hat q_i\hat p_j - \hat p_j \hat q_i=i\delta_{ij}$ 
(we set $\hbar =1$), which are summarized as 
\begin{equation}
\label{CCR}
   \hat x \hat x ^\top -(\hat x \hat x^\top )^\top 
       = i \Sigma_n,
\end{equation}
where $\Sigma_n$ is the following $2n\times 2n$ block diagonal matrix: 
\[
   \Sigma_n={\rm diag}\{\sigma, \ldots, \sigma\},~~
   \sigma = \left[\begin{array}{cc}
               0 & 1 \\
               -1 & 0
            \end{array}\right].
\]
The system variables are governed by the linear dynamics 
\begin{align}
\label{dynamics}
     d\hat x_t 
       =A\hat x_t dt+Fu_tdt + B d\hat W_t. 
\end{align}
Note that, in order to preserve Eq.~\eqref{CCR} for all $t$, the matrix $A$ 
must be of the following form:
\begin{equation}
\label{general A matrix}
    A=\Sigma_n(G+\Sigma_n^\top B \Sigma_m B^\top \Sigma_n/2), 
\end{equation}
where $G$ is a $2n\times 2n$ real symmetric matrix determining the system 
Hamiltonian by $\hat H=\hat x^\top G\hat x/2$. 
Also $B$ is a $2n \times 2m$ real matrix determined from the system-field 
coupling. 
$F$ is a real matrix and $u_t$ is the vector of classical (i.e., non-quantum) 
signal representing the control input. 
The system couples with $m$ probe or environment bosonic fields, with 
vector of noise operators 
$\hat W_t :=[\hat Q_1, \hat P_1, \ldots, \hat Q_m, \hat P_m]^\top$. 
This satisfies the quantum Ito rule $d\hat W_t d\hat W_t^\top=\Theta dt$, 
with zero mean: $\mean{d\hat W_t}=0$. 
The correlation matrix $\Theta$ is $2m\times 2m$ block diagonal 
Hermitian, and their $j$th block matrix (i.e., the correlation matrix of 
$\hat Q_j$ and $\hat P_j$) is in general written as 
\begin{equation}
\label{quantum Ito rule}
    \Theta_j = 
          \left[\begin{array}{cc}
               N_j+\Re(M_j)+1/2 & \Im(M_j)+i/2 \\
                 \Im(M_j)-i/2 & N_j-\Re(M_j)+1/2 \\
            \end{array}\right].
\end{equation}
The parameters $N_j\in{\mathbb R}$ and $M_j\in{\mathbb C}$ satisfy 
$N_j(N_j+1)\geq |M_j|^2$. 
Note that $N_j$ represents the average excitation number of the probe 
quanta, and $M_j$ is related to squeezing of the field; 
if $N_j(N_j+1)=|M_j|^2$ is satisfied, the probe field state is a pure squeezed state
\footnote{
The field annihilation operator $\hat A_1=(\hat Q_1 + i\hat P_1)/\sqrt{2}$ for 
a pure squeeze state is modeled by the Bogoliubov transformation 
$\hat A_1=\hat A_1^{(0)} \cosh(r/2) 
- \hat A_1^{(0)}\mbox{}^\dagger e^{i\theta/2}\sinh(r/2)$, where 
$\hat A_1^{(0)}$ is the vacuum field operator. 
The corresponding parameters $N_1$ and $M_1$ are obtained from the 
definition $d\hat A_1 d\hat A_1^\dagger =(N_1+1)dt$, 
$d\hat A_1^\dagger d\hat A_1=N_1dt$, 
$dA_1^2=M_1dt$, and $dA_1^\dagger\mbox{}^2=M_1^*dt$, which lead to 
$N_1=\sinh^2(r/2)$ and $M_1=-e^{i\theta/2}\sinh(r/2)\cosh(r/2)$. 
Hence certainly $N_1(N_1+1)=|M_1|^2$ is satisfied. 
}
, while if $M_j = 0$ it is not squeezed. 
Also note that $d\hat Q_j d\hat P_j - d\hat P_j d\hat Q_j=idt$.

For this system we perform a (joint) Homodyne measurement on $\ell~(\leq m)$ 
output probe fields, which generates the classical measurement signal 
\begin{align}
\label{output}
     d y_t = C\hat x_t dt + D d\hat W_t.
\end{align}
Note that, due to the unitarity of the system-field coupling, the $\ell\times 2n$ 
real matrix $C$ and the $\ell\times 2m$ real matrix $D$ satisfy the following 
specific structure:
\begin{equation}
\label{general C and D matrices}
    C=D\Sigma_m B^\top \Sigma_n, ~~~D\Sigma_m D^\top=0. 
\end{equation}

In this paper we assume that $A$ is {\it Hurwitz}, meaning that the real 
parts of all the eigenvalues of $A$ are negative; 
hence in this case the mean of the system variables, $\mean{\hat x_t}$, 
which obeys the dynamics $d\mean{\hat x_t}/dt=A\mean{\hat x_t}$, 
converges to zero in the long time limit, i.e., $\mean{\hat x_t}\rightarrow 0$. 
Note that the opto-mechanical system studied in Section 4 is Hurwitz.


\subsection{Quantum Kalman filter}

Let us consider the situation where we want to perform a real-time estimate 
of the system variable $\hat x_t$ based on the measurement signal $y_t$. 
The solution is provided by the quantum filtering theory; 
that is, we can rigorously define the quantum conditional expectation 
$\pi(\hat x_t):={\mathbb E}(\hat x_t\,|\,{\cal Y}_t)$, where ${\cal Y}_t$ is 
the $\sigma$-algebra composed of the measurement signal 
$\{ y_s\,|\,0\leq s\leq t\}$. 
In fact the classical random variable $\pi(\hat x_t)$ is the least mean squared 
estimate of $\hat x_t$. 
The recursive equation updating $\pi(\hat x_t)$ is given by the  quantum 
Kalman filter \cite{Doherty1999a,Doherty1999b,Belavkin2008}: 
\begin{eqnarray}
\label{linear-filter}
& & \hspace*{-2em}
     d\pi(\hat x_t) = A\pi(\hat x_t)dt + Fu_tdt 
                      + K_t(dy_t  -C\pi(\hat x_t)dt),
\nonumber \\ & & \hspace*{-2em}
     K_t =(V_t C^\top + B\Re(\Theta)D^\top)(D\Re(\Theta) D^\top)^{-1},
\end{eqnarray}
where the initial condition is $\pi(\hat x_0)=\mean{\hat x_0}$ with 
$\mean{\bullet}$ the unconditional expectation. 
$V_t$ is the estimation error covariance matrix defined as 
\[
    V_t:=\mean{
           \Delta\hat{x}_t\Delta\hat{x}_t^\top
         +(\Delta\hat{x}_t\Delta\hat{x}_t^\top)^\top}/2,~~
    \Delta\hat{x}_t:=\hat{x}_t-\pi(\hat x_t), 
\]
which evolves in time according to the following Riccati differential equation: 
\begin{equation}
\label{riccati}
\hspace{0.5 em}
     \dot V_t = AV_t + V_tA^\top + B\Re(\Theta)B^\top 
                        -K_t D\Re(\Theta) D^\top K_t^\top. 
\end{equation}
Under the assumption $A$ being Hurwitz
\footnote{
Note that the Hurwitz property is a sufficient condition for Eq.~\eqref{riccati} 
to have a unique steady solution. 
A useful necessary and sufficient condition is that $(A^\top, D)$ is stabilizable 
and $(A^\top, B)$ is detectable \cite{Kucera}. 
}
, this equation has a unique steady solution $V_\infty>0$.


\subsection{Quantum LQG control}

In the infinite horizon quantum LQG control problem, we consider the following 
cost function: 
\begin{equation}
\label{cost function}
    J[u]
       = \lim_{T\rightarrow \infty}
          \frac{1}{T}
          \biggl \langle
           \int_0^T ({\hat x}_t^\top Q {\hat x}_t
                    +u_t^{\top}Ru_t )dt \biggr \rangle, 
\end{equation}
where $Q\geq 0$ and $R>0$ are real weighting matrices. 
The goal is to design the feedback control law $u_t$ as a function of $y_t$, 
that minimizes the cost \eqref{cost function} under the condition \eqref{dynamics}, 
i.e., $u_t^*={\rm arg}\min_u J[u]$. 
The point is that, due to the tower property 
$\mean{\hat x_t}=\mean{\pi(\hat{x}_t)}
=\mean{{\mathbb E}(\hat x_t\,|\,{\cal Y}_t)}$, the cost can be represented 
in terms of only the filter variable as follows. 
That is, due to the relation 
${\mathbb E}(\hat{x}_t\hat{x}_t^\top \,|\, {\cal Y}_t)
=V_t + i\Sigma_n/2 + \pi(\hat{x}_t)\pi(\hat{x}_t)^\top$, we have 
\begin{eqnarray*}
& & \hspace*{0em}
      \biggl \langle 
                         \int_0^T (\hat{x}_t^\top Q \hat{x}_t + u_t^\top R u_t )dt 
                             \biggr \rangle
                = \biggl \langle 
                         \int_0^T 
                             \Big( \Tr[Q{\mathbb E}(\hat{x}_t\hat{x}_t^\top\,|\,{\cal Y}_t)]
                       + u_t^\top R u_t \Big)dt 
                             \biggr \rangle
\nonumber \\ & & \hspace*{3em}
       = \biggl \langle
           \int_0^T \Big( \pi(\hat{x}_t)^\top Q \pi(\hat{x}_t) 
                       + u_t^\top Ru_t \Big)dt \biggr \rangle
            + \int_0^T\Tr\Big[Q\Big(V_t+\frac{i}{2}\Sigma_n\Big)\Big]dt,
\end{eqnarray*}
where we have used the fact that $V_t$ obeys the deterministic time evolution 
\eqref{riccati}. 
Hence this equation leads to 
\begin{equation}
\label{cost function rewritten}
    J[u]
      = \lim_{T\rightarrow \infty}
          \frac{1}{T}
         \biggl \langle
           \int_0^T \Big( \pi(\hat{x}_t)^\top Q \pi(\hat{x}_t) 
                       + u_t^\top Ru_t \Big)dt \biggr \rangle
           + \Tr(QV_\infty).
\end{equation}
Note that the second term is constant. 
As a result, our problem is to find $u_t$ minimizing the first term of 
Eq.~\eqref{cost function rewritten} under the condition \eqref{linear-filter}. 
This is exactly the {\it classical} LQG control problem and can be analytically 
solved as follows (see \cite{Bensoussan} or Appendix~A); 
the optimal control input is given by 
\begin{equation}
\label{optimal control}
     u^*_t=-R^{-1}F^\top P_t\pi(\hat{x}_t),
\end{equation}
where $P_t$ is the solution of the following Riccati equation: 
\begin{equation}
\label{control Riccati}
     \dot{P}_t+P_tA + A^\top P_t - P_tFR^{-1}F^\top P_t + Q = 0. 
\end{equation}
Likewise the case of Eq.~\eqref{riccati}, because $A$ is Hurwitz
\footnote{
As mentioned in the footnote 2, this condition is stronger than the condition 
$(A,F)$ being stabilizable and $(A, \sqrt{Q})$ begin detectable, which is a 
necessary and sufficient condition for Eq.~\eqref{control Riccati} to have a 
unique steady solution $P_\infty\geq 0$. 
Note that, in this case, $A-FR^{-1}F^\top P_\infty$ is Hurwitz, meaning that 
the controlled filter equation is stable. 
}, this equation has a unique steady solution $P_\infty\geq 0$. 
The minimum value of the cost, which is reached by the optimal 
control \eqref{optimal control}, is given by 
\begin{equation}
\label{min cost}
    J[u^*]
       = \Tr(K_\infty D\Re(\Theta) D^\top K_\infty^\top P_\infty)
         + \Tr(QV_\infty).
\end{equation}
Note that $u^*_t$ is a function of the optimal estimate $\pi(\hat x_t)$. 
Thus, we can design the optimal estimate and control separately; 
this is called the {\it separation principle} as in the classical case 
\cite{Bouten 2008}.


\subsection{Lower bound of the minimum cost: The cheap control}

Let us set $R=\epsilon^2 I$ for the cost function \eqref{cost function}, where 
$\epsilon>0$ is a positive scalar, and use $P_\epsilon$ to denote the solution 
of the Riccati equation \eqref{control Riccati}. 
Then, it was proven in \cite{Sivan} that $P_\epsilon$ monotonically 
decreases as $\epsilon$ goes to zero. 
Moreover, we have that $\lim_{\epsilon\rightarrow 0}P_\epsilon=0$, if and 
only if the system characterized by $(A, F, \bar{Q})$ is {\it minimum phase} 
and {\it right invertible}; see Appendix~B for the definitions of this condition 
($\bar{Q}$ is a real matrix satisfying $Q=\bar{Q}^\top \bar{Q}$). 
In this case, the steady solution of the Riccati equation \eqref{control Riccati} 
takes the form $P_\infty=\epsilon \bar{P}+O(\epsilon^2)$. 
Then, the minimum cost \eqref{min cost} becomes 
\begin{equation}
\label{cheap cost}
     J[u^*] 
        = \epsilon \Tr(K_\infty D\Re(\Theta) D^\top K_\infty^\top \bar{P})
            + \Tr(QV_\infty) + O(\epsilon^2), 
\end{equation}
and the optimal control input \eqref{optimal control} is given by 
$u^*_t=-\epsilon^{-1}F^\top \bar{P}\pi(\hat{x}_t)$ at steady state. 
Now we consider the situation where the actuator is allowed to have a large 
control gain (e.g., the case $\epsilon\approx 0$); 
in particular, the control input in the ideal limit $\epsilon \rightarrow +0$, 
meaning that no penalty is imposed on it, is called the cheap control 
\cite{Sivan,Seron Book,Seron}. 
We then see that this ultimate feedback control perfectly suppresses the 
fluctuation of the estimated variables $\pi(\hat{x}_t)$, i.e., the first term 
of Eq.~\eqref{cheap cost}, and as a result the total cost is limited only by the 
optimal estimation error. 
Therefore, we can take $J^*[u^*] = \Tr(QV_\infty)$ as a fundamental 
quantity that is reasonably used for evaluating the performance of feedback 
control, because it cannot be further decreased by any control. 
In particular, we define the SQL as the value of $J^*[u^*]$ when the probe 
input is given by a coherent or a vacuum field. 
Finally note that, when $\epsilon\rightarrow +0$, Eq.~\eqref{cost function} 
equals to the stationary energy $\mean{\hat x_\infty^\top Q \hat x_\infty}$, 
and this can be ultimately reduced, by the ideal cheap control, to 
$J^*[u^*] = \Tr(QV_\infty)$.


\section{Configuration of the entanglement-assisted feedback control}


\subsection{Model}

The entanglement-assisted feedback control configuration considered in 
this paper is depicted in Fig.~\ref{General setup}. 
This model has the following four features. 

{\bf (i)} The system is linear and couples with a single probe field.

{\bf (ii)} 
The entangled optical field is produced by combining a fixed squeezed 
field $\hat W_1$ and a fixed coherent field $\hat W_2$, at a beam splitter (BS1). 
The correlation matrices \eqref{quantum Ito rule} of these fields are respectively 
given by 
\[
        \Theta_1 = \frac{1}{2}
                \left[ \begin{array}{cc}
                   \cos\theta & -\sin\theta \\
                   \sin\theta & \cos\theta \\
                \end{array} \right]
                \left[ \begin{array}{cc}
                   e^{-r} & i \\
                   -i & e^r \\
                \end{array} \right]
                \left[ \begin{array}{cc}
                   \cos\theta & \sin\theta \\
                   -\sin\theta & \cos\theta \\
                \end{array} \right],~~~
       \Theta_2 = \frac{1}{2}
                \left[ \begin{array}{cc}
                   1 & i \\
                   -i & 1 \\
                \end{array} \right],
\]
where $r$ is the squeezing level and $\theta$ represents the phase of the 
squeezed state in the phase space; see Fig.~\ref{General setup} and the footnote 
in page 5. 
The reflectivity of BS1 is, for simplicity, set to $\beta_1^2$ with 
$0\leq \beta_1\leq 1$. 
Note that, for all $\beta_1\in(0,1)$, the output fields of BS1 are entangled 
(see Appendix~C). 
As shown in the figure, one portion of this entangled field couples with 
the system, while the other portion does not. 
The degree of entanglement can be changed by tuning $\beta_1$, while 
maintaining the total amount of energy for producing this entangled field. 
Hence we can conduct a fair comparison of the entanglement-assisted 
control method (the case $0< \beta_1< 1$) and the standard method without 
entanglement (the case $\beta_1=0, 1$), given the same amount of resources. 
In particular, note that the SQL corresponds to the case $\beta_1=1$.

{\bf (iii)} We assume that the system's output field is subjected to an optical loss, 
which can be modeled by introducing a fictitious beam splitter with reflectivity 
$\delta^2$; if $\delta=0$, then there is no optical loss. 
$\hat W_3$ denotes the vacuum noise field coming into this fictitious beam 
splitter, whose correlation matrix is the same as that of $\hat W_2$, i.e., 
$\Theta_3 = \Theta_2$. 
As consequence the overall system contains three input fields $\hat W_1$, 
$\hat W_2$, and $\hat W_3$, implying that $m=3$ in the system equations 
\eqref{dynamics} and \eqref{output}. 
The total correlation matrix is thus given by 
\[
        \Theta={\rm diag}\{\Theta_1, \Theta_2, \Theta_3\}. 
\]
Note again that $\hat W_1$ and $\hat W_2$ represent the probe fields, 
while $\hat W_3$ denotes the unwanted noise field.

{\bf (iv)} The system's output field after being subjected to the optical loss meets 
the other portion of the entangled input field, at the second beam splitter 
(BS2) with reflectivity $\beta_2^2$; again for simplicity we assume 
$0\leq \beta_2 \leq 1$. 
Then the final output fields are measured by two Homodyne detectors with 
phase $\phi_1$ and $\phi_2$, which generate two output signals 
$y_1$ and $y_2$. 
Note that, if $\beta_2=0$ or $\beta_2=1$, the two optical fields are not 
combined at BS2 and are measured independently; this type of measurement 
is called the local measurement. 
The other case with $0<\beta_2<1$ is called the global measurement.

The overall system dynamics realizing the above setup is given as follows. 
First we use the fact that, for a general open linear system interacting with 
a single probe field, the system-field coupling is represented by an operator 
(the so-called Lindblad operator) of the form $\hat L=c^\top\hat x$ with 
$c$ the $2n$-dimensional complex column vector, and this determines the 
$B$ matrix in Eq.~\eqref{dynamics} as follows 
(see e.g., \cite{WisemanBook,KurtBook,Yamamoto2014}). 
That is, by defining the $2\times 2n$ real matrix 
\begin{equation}
\label{def of Cs}
     \bar{C}=
         \sqrt{2}\left[ \begin{array}{cc}
                         \Re(c)^\top  \\
                         \Im(c)^\top  \\
                      \end{array} \right], 
\end{equation}
we can specify the $B$ matrix in the following form:
\begin{equation}
\label{B matrix}
      B = [ \alpha_1\Sigma_n\bar{C}^\top\Sigma_1, 
               \beta_1\Sigma_n\bar{C}^\top\Sigma_1, 
               0_{2n\times 2} ],
\end{equation}
where $\alpha_1=\sqrt{1-\beta_1^2}$. 
Then the $A$ matrix is determined by Eq.~\eqref{general A matrix} 
with $G$ specified by the system Hamiltonian $\hat H=\hat x^\top G \hat x/2$. 
The $C$ matrix is also specified by Eq.~\eqref{general C and D matrices} and 
is now given by 
\begin{equation}
\label{C matrix}
     C = D\Sigma_3 B^\top \Sigma_n
        = D \left[ \begin{array}{c}
                   \alpha_1 \bar{C} \\
                   \beta_1 \bar{C} \\
                   0_{2\times 2n} \\
                \end{array} \right].
\end{equation}
Here $D$ is a $2\times 6$ real matrix of the form 
\[
     D = [D_1, D_2, O_2]T_2 T_L T_1,
\]
where 
\[
    D_1 = \left[ \begin{array}{cc}
                   \cos\phi_1 & \sin\phi_1 \\
                   0 & 0 \\
               \end{array} \right],~~~
     D_2 = \left[ \begin{array}{cc}
                   0 & 0 \\
                   \cos\phi_2 & \sin\phi_2 \\
               \end{array} \right]
\]
and 
\[
       T_k=\left[ \begin{array}{ccc}
                   \alpha_k I_2& \beta_k I_2 & O_2 \\
                   -\beta_k I_2& \alpha_k I_2 & O_2 \\
                   O_2 & O_2 & I_2 \\
                \end{array} \right]~~(k=1,2),~~~
       T_L=\left[ \begin{array}{ccc}
                   \sqrt{1-\delta^2} I_2 & O_2 & \delta I_2  \\
                   O_2 & I_2 & O_2 \\
                   -\delta I_2 & O_2 & \sqrt{1-\delta^2} I_2  \\
                \end{array} \right],
\]
with $\alpha_2=\sqrt{1-\beta_2^2}$. 
Note that $T_1$ and $T_2$ represent the scattering process at BS1 and BS2, 
respectively. 
Also $T_L$ corresponds to the optical loss in the system's output field. 
$D_1$ and $D_2$ represent the Homodyne measurements with phase 
$\phi_1$ and $\phi_2$, respectively.


\subsection{Information gain via entanglement}

This subsection is devoted to show that, in a special setup, an additional 
information about the system is indeed obtained through the second path 
in the interferometer, which may improve the control performance. 
That is, we consider the case where the system's output field completely 
diminishes, i.e., $\delta=1$. 
In this case, the $D$ matrix is given by 
\[
     D=[-\beta_1(\beta_2 D_1+\alpha_2 D_2), ~
              \alpha_1(\beta_2 D_1+\alpha_2 D_2), ~ \alpha_2 D_1-\beta_2 D_2], 
\]
and as a result $C=0$ for any choice of $\bar{C}$. 
Then the system equations \eqref{dynamics} and \eqref{output} are given by 
\begin{align}
\label{no knowledge dynamics}
     d\hat x_t 
       =A\hat x_t dt+Fu_tdt + B d\hat W_t, ~~
     d y_t = D d\hat W_t.
\end{align}
Hence, as expected, the measurement output $y_t$ does not {\it explicitly} 
contain any information about the system. 
However, interestingly, the observer can {\it implicitly} gain information; 
intuitively, this is because the observer knows that the {\it same} noise 
$\hat{W}_t$ enters into the system and the detector; 
in other words, the observer exactly knows the noise that drives the system and 
thus can track the estimate of the system's time-evolution. 
Actually, the quantum Kalman filter equation \eqref{linear-filter} is now 
given by 
\begin{equation}
\label{no knowledge filter}
     d\pi(\hat x_t) = A\pi(\hat x_t)dt + Fu_tdt 
                      + B\Re(\Theta)D^\top (D\Re(\Theta) D^\top)^{-1}dy_t, 
\end{equation}
which means that the observer can update the estimate $\pi(\hat x_t)$ 
using the measurement result $y_t$. 
Also the estimation error covariance matrix follows 
\begin{equation}
\label{no knowledge riccati}
     \dot V_t = AV_t + V_tA^\top + B\Re(\Theta)B^\top 
           -B\Re(\Theta)D^\top (D\Re(\Theta) D^\top)^{-1} D\Re(\Theta)B^\top. 
\end{equation}
Now, because $A$ is Hurwitz, $V_t$ has a steady solution, meaning that 
the estimation error is bounded. 
The above two equations indicate that the important term bringing the information 
to the filter \eqref{no knowledge filter} and \eqref{no knowledge riccati} 
is $B\Re(\Theta)D^\top$, which is now calculated as 
\[
      B\Re(\Theta)D^\top
       = \frac{\alpha_1\beta_1}{2}\Sigma_n \bar{C}^\top \Sigma_1
            [I_2-2\Re(\Theta_1)]
       \Big( \beta_2 D_1^\top + \alpha_2 D_2^\top \Big).
\]
If there is no entanglement (i.e., $r=0$ or $\alpha_1\beta_1=0$), then 
$B\Re(\Theta)D^\top=0$, and the filter equations are reduced to 
\[
     d\pi(\hat x_t) = A\pi(\hat x_t)dt + Fu_tdt, ~~~
     \dot V_t = AV_t + V_tA^\top + B\Re(\Theta)B^\top. 
\]
These are simply the dynamics of unconditional expectation 
$\pi(\hat x_t)=\mean{\hat x_t}$ and the error covariance matrix, 
which correspond to the master equation describing the statistical 
time-evolution of the system without measurement. 
Therefore it is now clear that the entanglement-assisted filter gains additional 
information about the system through the entangled input field. 
However, note that an additional information does not always improve the 
control performance, because, as demonstrated in 
Section \ref{The case of lossy output field}, an entangled probe field is generally 
fragile and as a result the system's output field becomes more noisy compared 
to the case of coherent input.

{\it Remark:} 
The measurement output $dy_t=Dd\hat{W}_t$ in 
Eq.~\eqref{no knowledge dynamics} has the form of 
{\it no-knowledge measurement} \cite{Szigeti}, which can be used to cancel 
decoherence. 
Interestingly, unlike the measurement scheme presented here, the no-knowledge 
one does not provide any information to the observer; 
actually for the setup of \cite{Szigeti} it can be proven that $A$ is not 
Hurwitz and the estimation error diverges.


\section{Entanglement-assisted feedback for opto-mechanical oscillator}

In this section, we conduct detailed numerical simulations to evaluate 
how much the proposed entanglement-assisted feedback control scheme 
is effective in a practical setup.


\subsection{System model}

\begin{figure}
\centering
\includegraphics[scale=0.36]{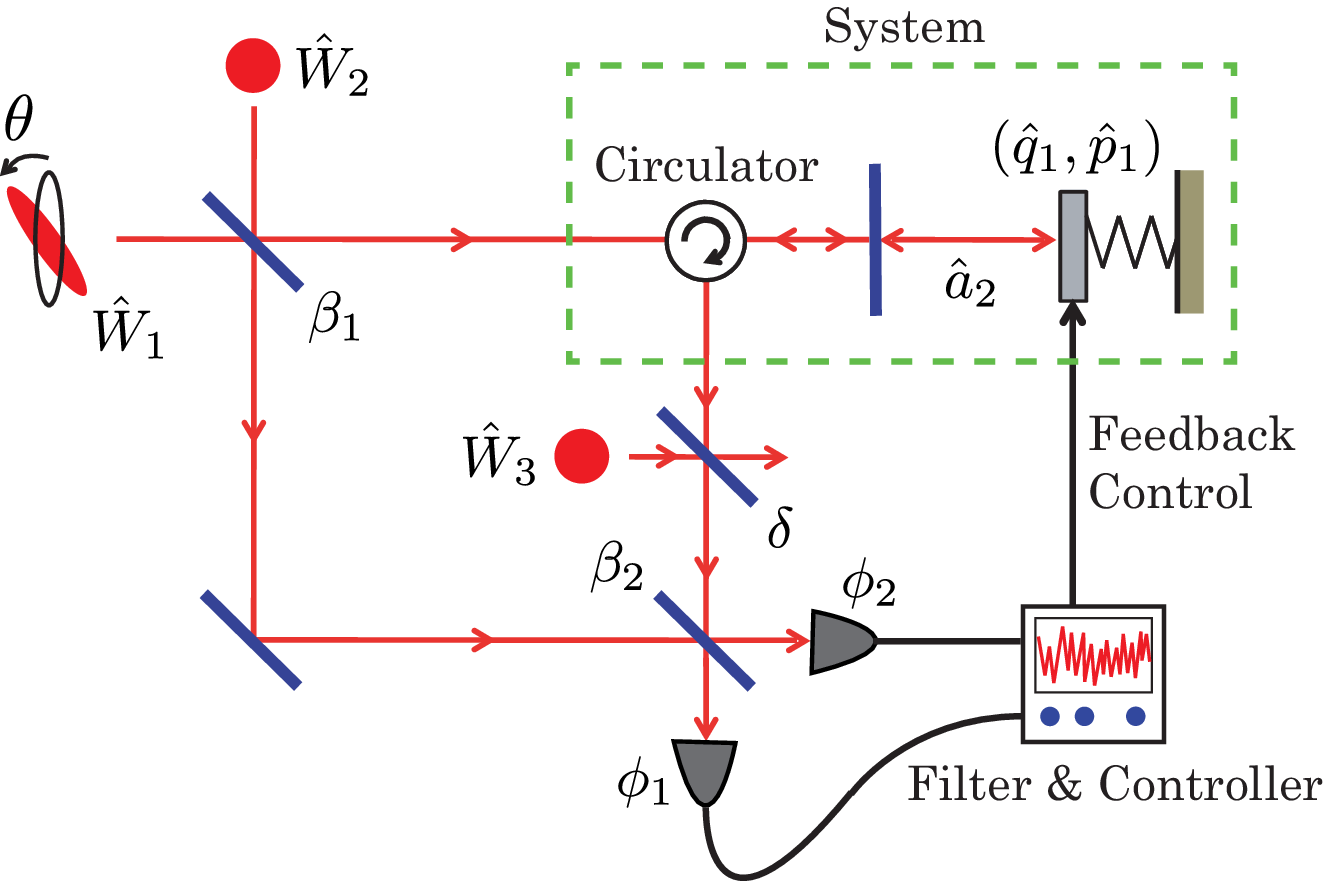}
\caption{\label{Setup}
Opto-mechanical oscillator coupled to the entangled probe field. }
\end{figure}

The system of interest is an opto-mechanical oscillator shown in Fig.~\ref{Setup}. 
Let $(\hat q_1, \hat p_1)$ be the position and momentum operators of the 
mechanical oscillator, and $\hat a_2$ be the annihilation operator of the optical 
cavity. 
The system Hamiltonian is given by 
\begin{equation}
\label{Osci Hamiltonian}
     \hat H = \frac{\omega}{2}(\hat q_1^2 + \hat p_1^2) 
                    + \frac{\Delta}{2}(\hat q_2^2 + \hat p_2^2) 
                       - \lambda \hat q_1 \hat q_2,
\end{equation}
where $\omega$ is the resonant frequency of the oscillator and $\Delta$ is the 
frequency detuning of the cavity mode in the rotating frame of the driving laser 
frequency; see \cite{Mancini 1998,Hofer 2015,Milburn} for more 
detailed description. 
Note $\hat q_2=(\hat a_2+\hat a_2^\dagger)/\sqrt{2}$ and 
$\hat p_2=(\hat a_2-\hat a_2^\dagger)/\sqrt{2}i$.  
The third term is the linearized radiation pressure force with strength 
$|\lambda|$, representing the interaction between the oscillator and 
the cavity field. 
From the relation $\hat H = \hat x^\top G\hat x/2$ with 
$\hat x = [\hat q_1, \hat p_1, \hat q_2, \hat p_2]^\top$, we have 
\[
    G = \left[ \begin{array}{cc|cc}
             \omega & 0 & -\lambda & 0 \\
             0 & \omega & 0 &  0 \\ \hline
             -\lambda & 0 & \Delta & 0 \\
             0 & 0 & 0 & \Delta \\
           \end{array} \right]. 
\]
The system couples to the driving laser  field at the partially reflective end-mirror 
of the cavity, with strength $\kappa$; 
this coupling is represented by the following operator: 
\[
     \hat L = \sqrt{\kappa} \hat a_2 
                 = \sqrt{\frac{\kappa}{2}}(\hat q_2 + i \hat p_2)
                 = \sqrt{\frac{\kappa}{2}}[0, 0, 1, i] \hat x. 
\]
Thus, Eq.~\eqref{def of Cs} yields 
\[
     \bar{C} 
         = \sqrt{\kappa}
         \left[ \begin{array}{cc|cc}
             0 & 0 & 1 & 0 \\
             0 & 0 & 0 & 1 \\
           \end{array} \right]. 
\]
This determines the system's $B$ and $C$ matrices from Eqs. \eqref{B matrix} 
and \eqref{C matrix}, respectively, and the $A$ matrix from 
Eq.~\eqref{general A matrix}. 
In addition, we assume that the oscillator is subjected to a thermal environment 
with mean photon number $\bar{n}_{\rm th}$. 
Then the system matrices are modified as follows; 
we need to change the $A$ matrix to $A-\gamma \Gamma/2$ and the constant 
term in the Riccati equation~\eqref{riccati}, $B\Re(\Theta)B^\top$, to 
$B\Re(\Theta)B^\top + \gamma(\bar{n}_{\rm th}+1/2)\Gamma$, where 
$\gamma$ represents the system-environment coupling strength and 
$\Gamma={\rm diag}\{1, 1, 0, 0\}$. 
Note that $A-\gamma \Gamma/2$ is Hurwitz, meaning that both the Riccati 
equations \eqref{riccati} and \eqref{control Riccati} have a unique steady 
solution. 
The oscillator can be directly controlled by implementing a piezo-actuator 
\cite{Rugar 2008} (the case shown in Fig.~\ref{Setup}) or indirectly 
controlled by modulating the input probe field. 
In both cases, it can be shown that the system satisfies the conditions for 
the cheap control described in Section 2.4; see Appendix~B.


\subsection{Control goal}

The control goal is to cool the oscillator toward its motional ground state; 
i.e., we want to minimize the stationary mechanical occupation number 
\[
       \bar{n} = \mean{\hat a_{1,\infty}^\dagger \hat a_{1,\infty}}
          = (\mean{\hat q_{1,\infty}^2}+\mean{\hat p_{1,\infty}^2}-1)/2. 
\]
As described in Section 2.4, this can be ultimately reduced, by the ideal cheap 
control, to 
\begin{equation}
\label{min energy}
      \bar{n}^*=(\Tr(\Gamma V_\infty)-1)/2,  
\end{equation}
where again $\Gamma={\rm diag}\{1, 1, 0, 0\}$ and $V_\infty$ is the 
steady solution of the Riccati equation \eqref{riccati}.

The system parameters are set to the following typical values (in the unit 
$\omega=1$) in the feedback cooling setup (e.g., \cite{Hofer 2015}). 
First we assume the resonant driving $\Delta=0$, meaning that the oscillator's 
position and momentum can be best estimated by the filter and accordingly 
controlled efficiently. 
Also the cavity line width is set to $\kappa=2$ (bad cavity regime), so that 
the intra cavity field immediately leaks to outside and as a consequence 
the oscillator dynamics can be well observed by the filter. 
The oscillator is subjected to a thermal noise with mean photon number 
$\bar{n}_{\rm th}=1\times 10^5$ with coupling strength 
$\gamma=1\times 10^{-7}$. 
The squeezing level of the probe field is set to $r=2.3$ (10 dB squeezing), 
which is accessible with the current technology.

In this setting, the task is to optimize the parameters 
$(\beta_1, \beta_2, \phi_1, \phi_2, \theta)$ so that Eq.~\eqref{min energy} 
is minimized. 
To see how to find those optimal parameters, let us assume the lossless setup 
(i.e., $\delta=0$) and focus on only the oscillator mode, where the cavity mode is adiabatically eliminated. 
This dynamical equation is obtained by setting $d\hat q_2=0$ and 
$d\hat p_2=0$ due to $\kappa\gg \gamma$ and eventually eliminating 
$(\hat q_2, \hat p_2)$ from the whole dynamical equation; 
\begin{eqnarray}
& & \hspace*{-1em}
\label{q1 eq}
     d\hat{q}_1 = -\frac{\gamma}{2} \hat q_1 dt + \omega \hat p_1 dt 
                           -\sqrt{\gamma}d\hat Q_{\rm th}, 
\\ & & \hspace*{-1em}
\label{p1 eq}
     d\hat{p}_1 = -\omega\hat q_1 dt -\frac{\gamma}{2} \hat p_1 dt + udt 
                            -\frac{2\lambda}{\sqrt{\kappa}}
                                       (\alpha_1 d\hat Q_1 + \beta_1 d\hat Q_2)
                           -\sqrt{\gamma}d\hat P_{\rm th}, 
\\ & & \hspace*{-1em}
\label{y1 eq}
      dy_1 = \frac{2\alpha_2\lambda \sin\phi_1}{\sqrt{\kappa}} \hat q_1 dt
                  - (\alpha_1 \alpha_2 + \beta_1\beta_2 ) 
                       ( \cos\phi_1 d\hat Q_1  +  \sin\phi_1 d\hat P_1)
\nonumber \\ & & \hspace*{8.6em}
                \mbox{}
                  + (\alpha_1 \beta_2 - \beta_1\alpha_2 ) 
                       ( \cos\phi_1 d\hat Q_2  +  \sin\phi_1 d\hat P_2 ),
\\ & & \hspace*{-1em}
\label{y2 eq}
      dy_2 = -\frac{2\beta_2\lambda \sin\phi_2}{\sqrt{\kappa}} \hat q_1 dt
                  + (\alpha_1 \beta_2 - \beta_1\alpha_2 ) 
                       ( \cos\phi_2 d\hat Q_1  +  \sin\phi_2 d\hat P_1 )
\nonumber \\ & & \hspace*{9.3em}
                \mbox{}
                  + (\alpha_1 \alpha_2 + \beta_1\beta_2 ) 
                       (\cos\phi_2 d\hat Q_2 + \sin\phi_2 d\hat P_2 ),
\end{eqnarray}
where $u$ represents the magnitude of the force applied to a piezo-actuator 
mounted on the oscillator, and $(\hat Q_{\rm th}, \hat P_{\rm th})$ are the 
quadrature of the thermal field. 
These equations lead to a rough guide for choosing the parameters, as follows. 
\begin{enumerate}

\item
First, the bigger the first terms in Eqs.~\eqref{y1 eq} and \eqref{y2 eq} 
become, the more information the observer gains. 
Hence, it would be reasonable to make the terms $\sin\phi_1$ and $\sin\phi_2$ 
bigger, or equivalently take $\phi_1\approx \pi/2$ and $\phi_2\approx \pi/2$. 

\item
The above choice of $(\phi_1, \phi_2)$ implies that the measurement outputs 
are dominantly affected by the phase-quadrature noise $(\hat P_1, \hat P_2)$ 
rather than the amplitude quadrature noise $(\hat Q_1, \hat Q_2)$. 
Then by squeezing $\hat P_1$, we can improve the signal to noise ratio both 
in $y_1$ and $y_2$. 
This means that $\theta\approx \pi/2$ would be a proper choice. 

\item
The coefficients of the noise terms related to $\hat W_1=[\hat Q_1, \hat P_1]^\top$ 
and $\hat W_2=[\hat Q_2, \hat P_2]^\top$ in $y_1$ and $y_2$ cannot be 
simultaneously reduced, because they satisfy 
$(\alpha_1 \alpha_2 + \beta_1\beta_2 )^2 + 
(\alpha_1 \beta_2 - \beta_1\alpha_2 )^2=1$. 
Then, because $\hat W_2$ is not a tunable noise field, meaning that 
$\mean{(\cos\phi_1 d\hat Q_2  +  \sin\phi_1 d\hat P_2)^2}=dt$ and 
$\mean{(\cos\phi_2 d\hat Q_2  +  \sin\phi_2 d\hat P_2)^2}=dt$, it would 
be reasonable to choose the BS parameters so that the coefficient of $\hat W_2$ 
is reduced; 
more precisely, $\alpha_1 \beta_2 - \beta_1\alpha_2\approx 0$ if $y_1$ is 
mainly used (i.e., $\alpha_2\approx 1$), or 
$\alpha_1 \alpha_2 + \beta_1\beta_2\approx 0$ if $y_2$ is mainly used 
(i.e., $\beta_2\approx 1$). 
This leads to $(\alpha_1, \beta_1, \alpha_2, \beta_2)\approx (1, 0, 1, 0)$ or 
$(\alpha_1, \beta_1, \alpha_2, \beta_2)\approx (1, 0, 0, 1)$. 

\end{enumerate}
Of course, the above intuitive observation, particularly the last one, would not 
be necessarily true. 
In fact, we have now arrived at $(\alpha_1, \beta_1)\approx (1, 0)$, but this 
means that the input probe field is nearly a separable state where the squeezed 
component is injected to the system, and the entanglement property is not 
effectively used. 
Moreover, when $(\alpha_1, \beta_1)\approx (1, 0)$, the back-action noise on 
$\hat p_1$ (i.e., the fourth term in the right-hand side of Eq.~\eqref{p1 eq}) is 
dominated by $d\hat Q_1$; 
however, because now $\hat Q_1$ is nearly {\it anti-squeezed} 
(due to $\theta\approx \pi/2$), this parameter choice induces a bigger 
back-action noise. 
Therefore, the parameters have to be carefully chosen, via detailed numerical 
simulations taking into account the tradeoff between the back-action noise 
and the signal to noise ratio for the measurement outputs $y_1$ and $y_2$.


\subsection{Effectiveness of the entanglement-assisted feedback control}

\begin{figure}
\centering
\includegraphics[scale=0.45]{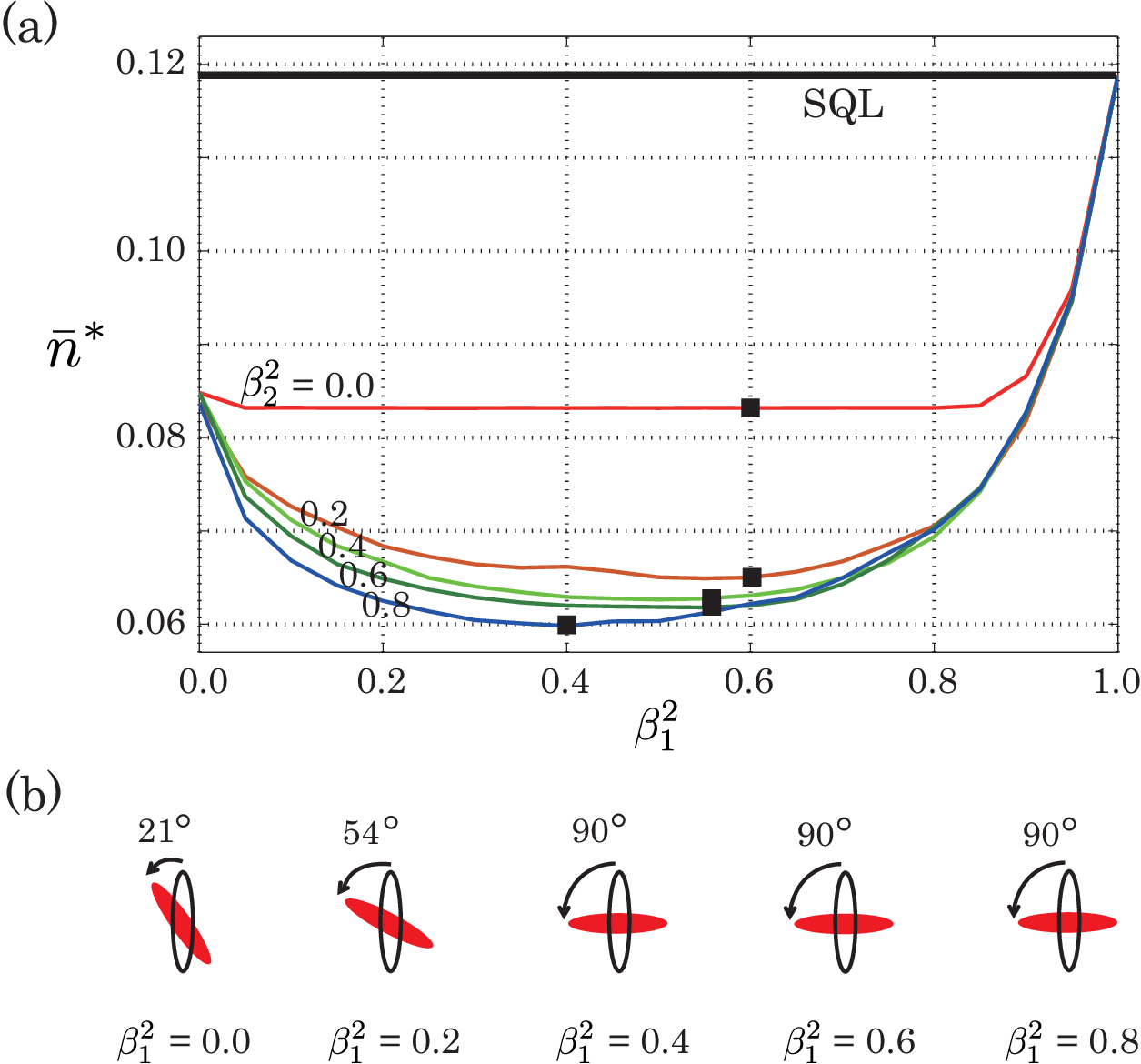}
\caption{\label{Cooling optim}
(a) The achievable lowest mechanical occupation number, $\bar{n}^*$, versus 
$\beta_1^2$ (the reflectivity of BS1) for several values of $\beta_2^2$, 
for the case $\lambda=0.3$ and $\delta=0$. 
At each point of $\beta_1$, the phase of the squeezed field and the Homodyne 
detection, $(\theta, \phi_1, \phi_2)$, are optimized. 
The black box indicates the minimum of $\bar{n}^*$. 
(b) The optimal phase $\theta$ of the squeezed field at each $\beta_1$ 
for the case $\beta_2^2=0.8$. 
}
\end{figure}

First let us see if the entanglement would actually bring any advantage 
to the feedback control. 
Figure~\ref{Cooling optim} (a) shows $\bar{n}^*$ as a function of the 
reflectivity of BS1, $\beta_1^2$, in the case $\lambda=0.3$ (weak coupling 
regime) and $\delta=0$ (the system's output field has no loss). 
$\bar{n}^*$ is calculated from Eq.~\eqref{min energy} together with the 
steady solution $V_\infty$ of the Riccati equation \eqref{riccati}. 
Furthermore, it is minimized with respect to the phase of the probe squeezed 
field, $\theta$, and the phases of the two Homodyne detectors, $(\phi_1,\phi_2)$, 
at each $\beta_1$; 
Figure~\ref{Cooling optim} (b) illustrates the optimal $\theta$ at each $\beta_1$ 
for the case $\beta_2^2=0.8$. 
The reflectivity $\beta_1^2$ represents how much the squeezed field $\hat W_1$ 
is split into two arms, which determines the amount of entanglement. 
Note that when $\beta_1=0$ or $\beta_1=1$, the input fields are not entangled. 
In particular, $\beta_1=1$ corresponds to the standard case where only 
the coherent field is injected to the system; 
hence the value of $\bar{n}^*$ in this case has the meaning of SQL, which is 
now $\bar{n}^*_{\rm SQL}\approx 0.119$ as indicated in the figure (a). 
The five solid curves in the figure (a) show $\bar{n}^*$ for several values of 
$\beta_2^2$, the reflectivity of BS2; 
recall that $\beta_2=0$ means the case of local measurement, while the 
cases $\beta_2\neq 0$ correspond to the global measurement (see {\bf (iv)} 
in Section 3.1). 
Importantly, in all cases the minimum of $\bar{n}^*$, which is indicated by the 
black box, is smaller than the SQL and is attained at a certain point of 
$\beta_1\in(0,1)$, where the input field is entangled. 
In particular, the most effective feedback cooling is carried out when we use 
the highly entangled probe field with $\beta_1^2=0.4$ and perform the 
global measurement with $\beta_2^2=0.8$, in which case the minimum of 
$\bar{n}^*$ is about 0.06. 
As a conclusion, the entanglement-assisted feedback control is in fact 
effective and realizes further cooling of the oscillator below the SQL. 
The followings are the list of other notable features of this system. 
\begin{itemize}

\item
For the cases $\beta_1^2=0.4, 0.6, 0.8$, the optimal phase of the squeezed 
field is $\theta=\pi/2$. 
The optimality of $\theta=\pi/2$ was indeed expected in the second observation 
in Section~4.2 (page 13), but $\beta_1$ is not nearly zero, which is not consistent 
with the third observation in Section~4.2. 
Therefore, the numerical solver has actually chosen a nontrivial set of parameters 
that balances the back-action noise on $\hat p_1$ and the measurement noise on 
$(y_1, y_2)$. 

\item
The minimum of $\bar{n}^*$ is reached at $\beta_1^2\neq 0.5$ and 
$\theta=\pi/2$. 
This means that the maximal entangled field is not the best probe for the 
estimation and feedback control; see Appendix~C. 

\item
The entanglement-assisted method outperforms the control with the optimized 
squeezed probe field \cite{Andersen 2016}, which corresponds to the case 
$\beta_1=0$. 

\end{itemize}
%


\subsection{Coupling strength and optimal probe}

\begin{figure}
\centering
\includegraphics[scale=0.39]{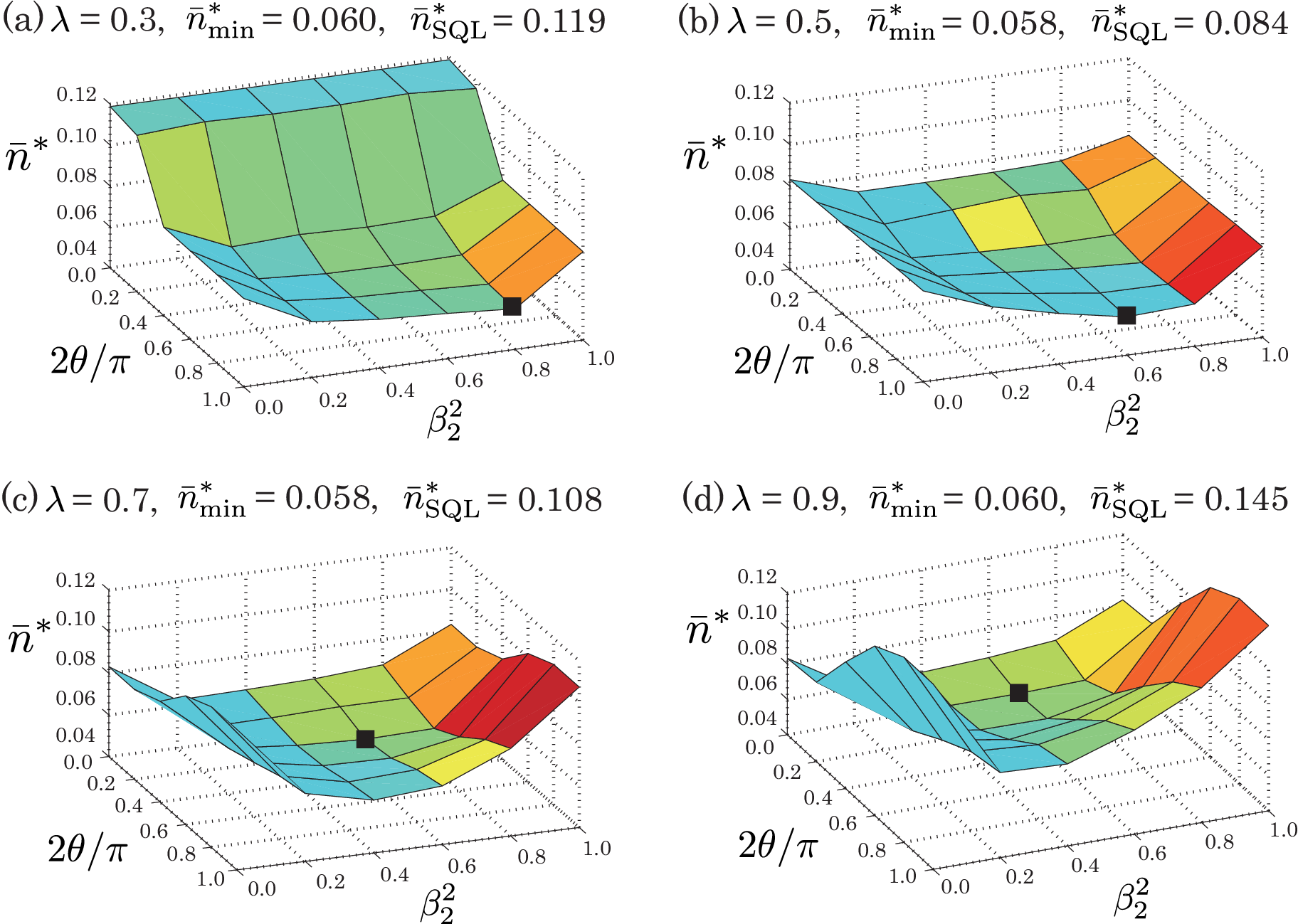}
\caption{\label{Cooling optim 3D}
The achievable lowest mechanical occupation number, $\bar{n}^*$, as a function 
of $\theta$ (phase of the squeezed field) and $\beta_2^2$ (reflectivity of 
BS2), for $\delta=0$ and several values of $\lambda$ (strength of the radiation 
pressure). 
The black box indicates the minimum of $\bar{n}^*$. 
}
\end{figure}

Here we study how much the minimum occupation number $\bar{n}^*_{\rm min}$ 
changes with respect to the coupling strength $\lambda$. 
Figure \ref{Cooling optim 3D} shows $\bar{n}^*$ as a function of $\theta$ 
and $\beta_2$, for $\delta=0$ and several values of $\lambda$. 
In each figure (a)-(d), $\bar{n}^*$ is already minimized with respect to 
$(\beta_1, \phi_1, \phi_2)$, and the achieved $\bar{n}^*_{\rm min}$ is 
shown together with $\bar{n}^*_{\rm SQL}$. 
In particular, in each figure, the optimal value of $\beta_1$ has been chosen 
as: (a) $\beta_1^2=0.40$, (b) $\beta_1^2=0.65$, (c) $\beta_1^2=0.55$, 
and (d) $\beta_1^2=0.50$, implying that the input probe field is highly 
entangled in all cases. 
Hence, we end up with the same conclusion that the entanglement-assisted 
feedback control cools the oscillator below the SQL and even performs better 
than the case with optimized squeezed probe field.

Note here that, as implied by Eqs.~\eqref{p1 eq}, \eqref{y1 eq}, and \eqref{y2 eq}, 
making $\lambda$ bigger improves the signal to noise ratio in the measurement 
output, but at the same time this induces a bigger back-action noise on $\hat p_1$. 
Hence, $\bar{n}^*_{\rm SQL}$ does not monotonically change with respect to 
$\lambda$; 
interestingly, $\bar{n}^*_{\rm min}$ takes almost the same value for all 
$\lambda$, which suggests that there would exist a fundamental lower 
bound of $\bar{n}^*_{\rm min}$ that is independent to $\lambda$. 
Another remarkable fact is that, for small values of $\lambda$, the optimal 
phase of the squeezed field is $\theta=\pi/2$, as seen in the previous 
subsection; however, this does not hold when $\lambda$ becomes large. 
This is because, for a large $\lambda$, it is more important to reduce 
the back-action noise $(2\lambda/\sqrt{\kappa})\alpha_1 d\hat Q_1$ 
than to improve the signal to noise ratio in the measurement process, and 
thus the squeezed field with $\mean{d\hat Q_1^2}<\mean{d\hat P_1^2}$ 
is chosen.


\subsection{The case of lossy output field}
\label{The case of lossy output field}

\begin{figure}
\centering
\includegraphics[scale=0.35]{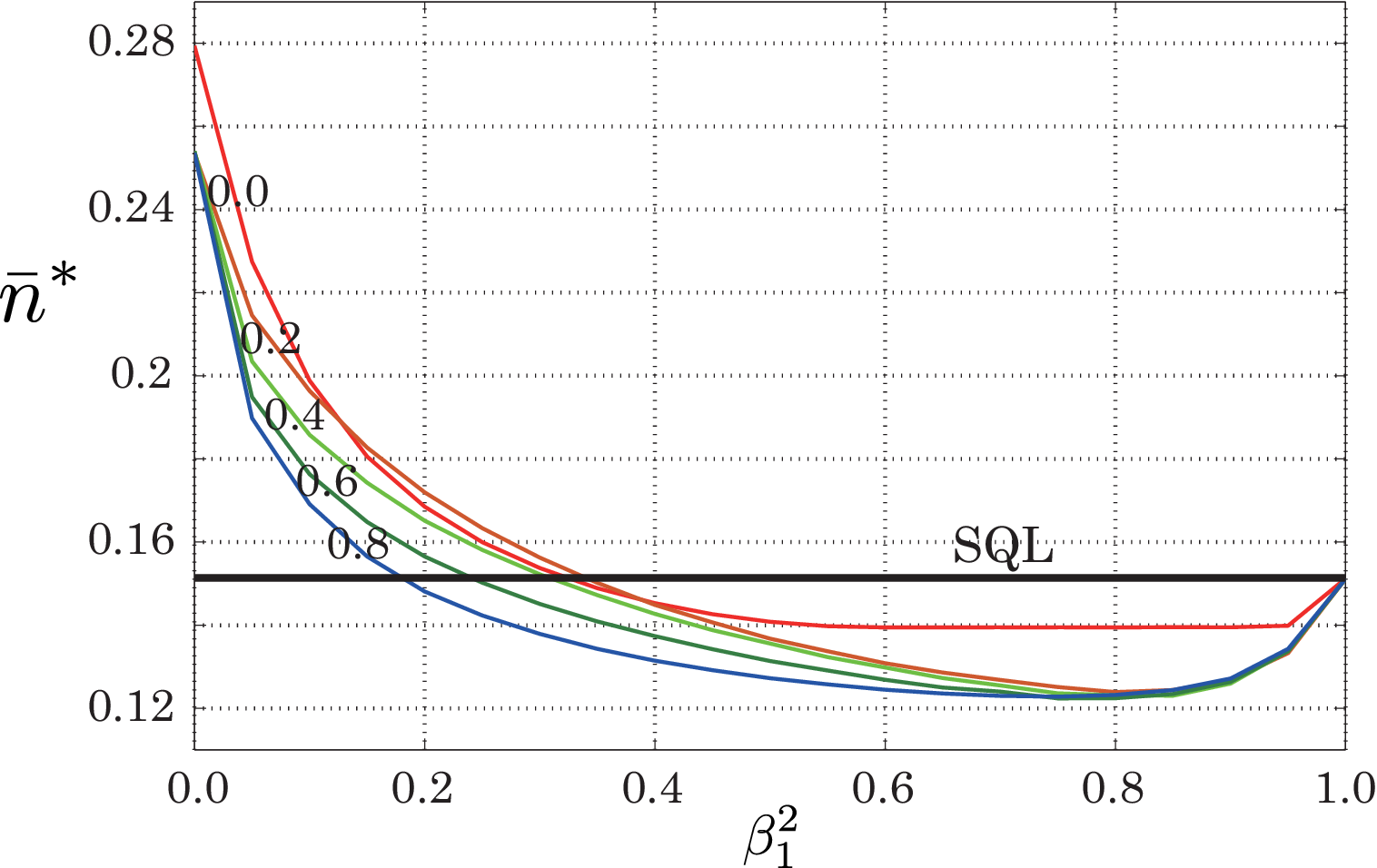}
\caption{\label{loss curves}
The achievable lowest mechanical occupation number, $\bar{n}^*$, versus 
$\beta_1^2$ (the reflectivity of BS1) for several values of $\beta_2^2$, for 
the case $\lambda=0.3$ and $\delta^2=0.1$. 
At each point of $\beta_1$, the phase of the squeezed light field and the 
Homodyne detection, $(\theta, \phi_1, \phi_2)$, are optimized. 
The numbers indicated along the curves are the values of $\beta_2^2$, as 
in the case of Fig.~\ref{Cooling optim}. 
}
\end{figure}

Next let us consider the case where the system's output field is subjected to 
the optical loss. 
Figure~\ref{loss curves} is the plot of $\bar{n}^*$ with the same setting as 
in Fig.~\ref{Cooling optim} (i.e., $\lambda=0.3$ and $(\theta, \phi_1, \phi_2)$ 
are optimized), except that the loss parameter is now set to $\delta^2=0.1$. 
As in the lossless case $\delta=0$, we find that the minimum of $\bar{n}^*$ 
is reached when the input probe field is entangled ($\beta_1^2\approx 0.8$) 
and the global measurement ($\beta_2^2=0.8$) is performed. 
However, notably, the difference between the minimum value of $\bar{n}^*$ 
and the SQL given at $\beta_1=1$ (i.e., how much the control performance is 
improved by entanglement) is smaller than the case when $\delta=0$. 
That is, the entanglement-assisted feedback is less effective if the system's 
output field is lossy. 
Another notable feature is that there is a case where the control performance 
becomes worse than the SQL via the entanglement-assisted feedback control. 
This happens when $\beta_1$ takes a small value, in which case 
the portion injected into the system is nearly a pure squeezed field. 
This result makes sense, because, as is well known, a squeezed field is fragile 
to noise and the system's output field loses more information than the 
standard case, which cannot be compensated by the additional information 
gained from the second path of the interferometer.

\begin{figure}
\centering
\includegraphics[scale=0.39]{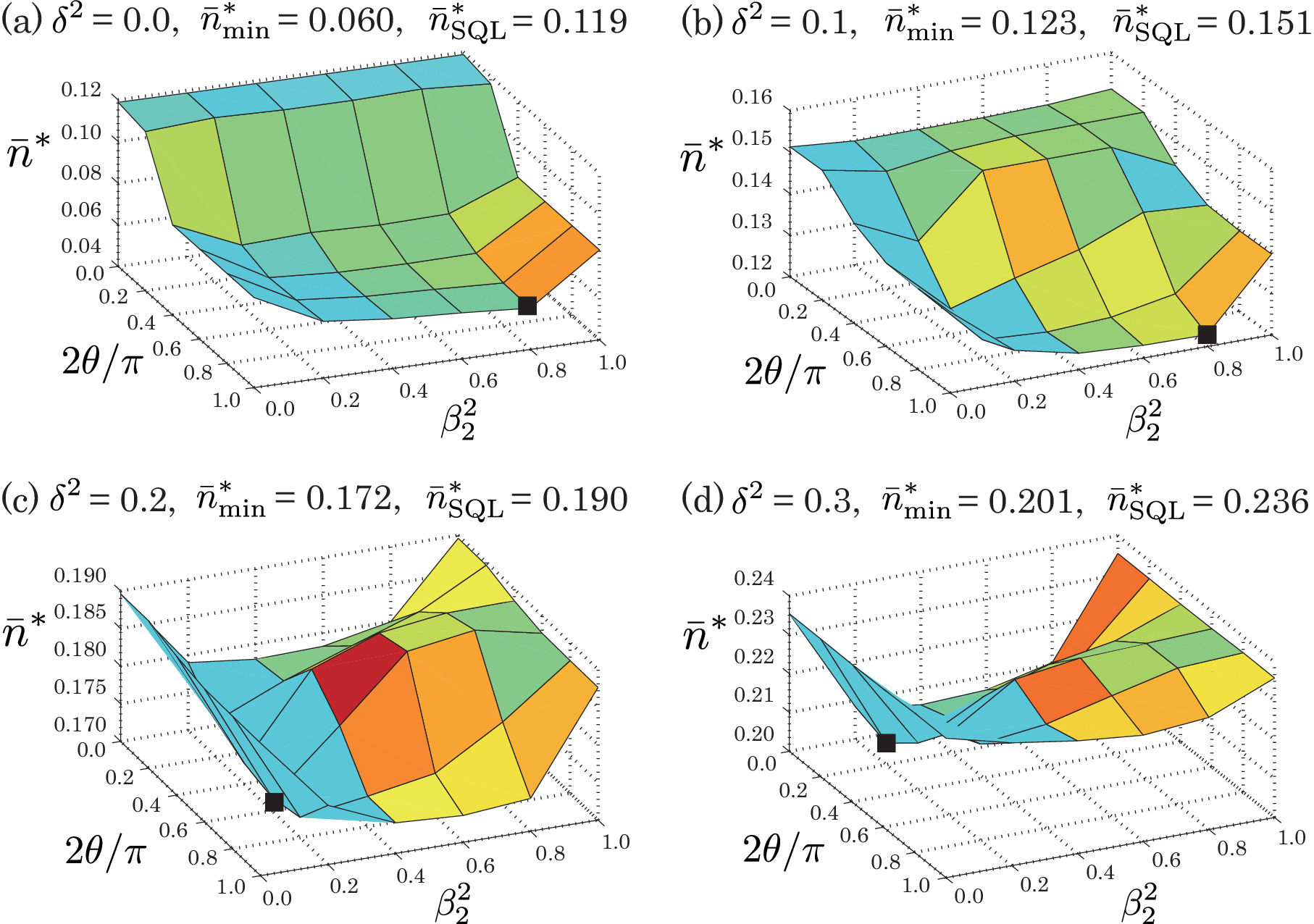}
\caption{\label{Cooling loss 3D}
The achievable lowest mechanical occupation number, $\bar{n}^*$, as a function 
of $\theta$ (phase of the squeezed light field) and $\beta_2^2$ 
(reflectivity of BS2), for $\lambda=0.3$ and several vales of $\delta$ 
(loss in the system's output field). 
The black box indicates the minimum of $\bar{n}^*$. 
}
\end{figure}

Finally Fig.~\ref{Cooling loss 3D} shows the plot of $\bar{n}^*$ as a function of 
$\theta$ and $\beta_2$, for $\lambda=0.3$ and several values of $\delta$. 
As in the case of Fig.~\ref{Cooling optim 3D}, $\bar{n}^*$ is already minimized 
with respect to $(\beta_1, \phi_1, \phi_2)$. 
Note that Fig.~\ref{Cooling loss 3D} (a) is the same as Fig.~\ref{Cooling optim 3D} 
(a). 
A notable point is that the optimal values of $\theta$ and $\beta_2$ in the case 
$\delta^2=0.1$ are the same as those for $\delta=0$. 
This means that the optimal input probe field and measurement are independent 
to the system's output loss $\delta$. 
This is a desirable fact because an exact value of $\delta$ is hard to estimate 
in practice, but the same input probe field and measurement can be used 
without respect to $\delta$ as long as the system's output loss is enough 
suppressed. 
However, Figs.~\ref{Cooling loss 3D} (c) and (d) show that the probe and 
measurement have to be changed when $\delta$ becomes bigger. 
In this sense, the optimal probe field is not robust for a system with lossy output.

\section{Conclusion}

In this paper, we have formulated the entanglement-assisted feedback control 
method for general linear quantum systems, which involves optimization of the 
amount of entanglement, the phase of the probe squeezed field, and the 
Homodyne measurement. 
Thanks to the linear setting, the strict lower bound of LQG cost function, which 
is achievable by the ideal cheap control, can be explicitly obtained, and it is 
used to evaluate the control performance. 
In the detailed numerical simulation studying the cooling problem of an 
opto-mechanical oscillator, it was shown that the entanglement-assisted controller 
works better than the standard method without entanglement, i.e., the control 
with a coherent probe field and even that with an optimized squeezed probe field. 
Although the improvement is not so drastic especially when the system's output 
field is lossy, we expect that a significant advantage of the entanglement-assisted 
method would appear for some nonlinear systems. 
In fact, it was shown in \cite{Clerk 2015} that, in a different measurement 
configuration, an entangled probe field can be used to significantly improve 
the detection efficiency for a qubit system; 
an extension of this study to the feedback control problem is an interesting 
future work.


\section*{Appendix A: Solution of LQG control problem}

Here we briefly explain how to derive the solution of the LQG control problem; 
see \cite{Bensoussan} for a more detailed derivation. 
The essential idea is to use the {\it dynamic programming} method based on 
the following {\it expected cost-to-go}: 
\[    
   J_t[u, z]
      ={\mathbb E}\Big[
            \int_t^T \Big( \pi(\hat{x}_s)^\top Q \pi(\hat{x}_s) 
                       + u_s^\top Ru_s \Big)ds ~\Big|~ \pi(\hat{x}_t)=z \Big]. 
\]
The goal is to obtain the minimum of this function, i.e., 
$J_t^*(z)=\min_{u[t,T]}J_t[u,z]$, with respect to the input in the time 
interval $[t, T]$, denoted by $u[t,T]$. 
Now we rewrite $J_t^*(z)$ in the following form:
\begin{eqnarray}
& & \hspace*{-1em}
   J_t^*(z)
      = \min_{u[t,T]}{\mathbb E}\Big[
            \int_t^{t+dt} \Big( \pi(\hat{x}_s)^\top Q \pi(\hat{x}_s) 
                       + u_s^\top Ru_s \Big)ds 
\nonumber \\ & & \hspace*{6em}
      \mbox{} 
             + \int_{t+dt}^T \Big( \pi(\hat{x}_s)^\top Q \pi(\hat{x}_s) 
                  + u_s^\top Ru_s \Big)ds              
                       ~\Big|~ \pi(\hat{x}_t)=z \Big] 
\nonumber \\ & & \hspace*{1.6em}
       = \min_{u_t}\Big\{
              (z^\top Q z 
                       + u_t^\top Ru_t )dt + J_{t+dt}^*(z+dz) \Big\}.
\nonumber
\end{eqnarray}
Then, noting that $\pi(\hat{x}_t)$ obeys Eq.~\eqref{linear-filter} and 
$d\bar{w}_t=dy_t  -C\pi(\hat x_t)dt$ is the standard classical Wiener 
process satisfying $d\bar{w}_td\bar{w}_t^\top =D\Theta D^\top dt$, 
we see that the optimal value function $J_t^*(z)$ satisfies the 
{\it Bellman equation} 
\begin{eqnarray}
& & \hspace*{-1em}
     \min_{u_t}\Big\{
       \Big\|u_t
       +\half R^{-1}F^\top\frac{\partial J_t^*(z)}{\partial z}\Big\|_R^2
        -\frac{1}{4}\Big(\frac{\partial J_t^*(z)}{\partial z}\Big)^\top
          FR^{-1}F^\top \frac{\partial J_t^*(z)}{\partial z}
        + \frac{\partial J_t^*(z)}{\partial t}
\nonumber \\ & & \hspace*{2em}
    \mbox{}
     + z^\top Q z 
     +\Big(\frac{\partial J_t^*(z)}{\partial z}\Big)^\top Az
     + \half \Tr\Big[ \frac{\partial^2 J_t^*(z)}{\partial z\mbox{}^2} 
                       K_t D \Re(\Theta) D^\top K_t^\top\Big] \Big\}=0, 
\nonumber
\end{eqnarray}
with the terminal condition $J_T^*(z)=0$. 
Here we have defined $\|x\|_R^2=x^\top Rx$. 
The optimal control input is thus given by 
\begin{equation}
\label{optimal-u-V}
    u_t^*=-\half R^{-1}F^\top\frac{\partial J_t^*(z)}{\partial z}. 
\end{equation}
The optimal value function $J_t^*(z)$ is then determined from the 
following partial differential equation: 
\begin{eqnarray}
& & \hspace*{-1em}
      \frac{\partial J_t^*(z)}{\partial t}
        +z^\top Q z
        -\frac{1}{4}\Big(\frac{\partial J_t^*(z)}{\partial z}\Big)^\top
          FR^{-1}F^\top \frac{\partial J_t^*(z)}{\partial z}
    +\Big(\frac{\partial J_t^*(z)}{\partial z}\Big)^\top Az
\nonumber \\ & & \hspace*{2.24em}
\mbox{}
    + \half \Tr\Big[ \frac{\partial^2 J_t^*(z)}{\partial z\mbox{}^2} 
                       K_t D \Re(\Theta) D^\top K_t^\top\Big]=0. 
\nonumber
\end{eqnarray}
Now we assume that the solution is of the quadratic form 
$J_t^*(z)=z^\top P_t z + \nu_t$ with $P_t\in{\mathbb R}^{2n\times 2n}$ 
and $\nu_t\in{\mathbb R}$; 
then the above partial differential equation is reduced to 
\begin{eqnarray}
& & \hspace*{-1em}
    z^{{\mathsf T}}\Big(
       \dot{P}_t+P_tA+A^\top P_t-P_t FR^{-1}F^\top P_t+Q \Big)z
\nonumber \\ & & \hspace*{2.24em}
\mbox{}
           + \dot{\nu}_t 
               + \Tr[ P_t K_t D \Re(\Theta) D^\top K_t^\top]=0. 
\nonumber
\end{eqnarray}
This equality must hold for any $z\in{\mathbb R}^{2n}$, and we thus obtain 
the following set of ordinary differential equations: 
\[
     \dot{P}_t+P_tA+A^\top P_t-P_t FR^{-1}F^\top P_t+Q=0, ~~
     \dot{\nu}_t + \Tr[ P_t K_t D \Re(\Theta) D^\top K_t^\top] = 0. 
\]
It follows from $J_T^*(z)=0$ that the terminal conditions are $P_T=0$ and 
$\nu_T=0$. 
Under the assumption that the above set of equations have solutions, 
the optimal controller (\ref{optimal-u-V}) is given by 
\begin{equation}
\label{optimal-u}
    u^*_t=-R^{-1}F^\top P_t\pi(\hat{x}_t). 
\end{equation}
Moreover, we now have
\[
     \mean{J_0^*(\hat{x}_0)} 
         = \mean{ \hat{x}_0^\top P_0 \hat{x}_0 + \nu_0}
         = \mean{\hat{x}_0^\top P_0 \hat{x}_0} 
              + \int_0^T \Tr[ P_t K_t D \Re(\Theta) D^\top K_t^\top]dt,
\]
hence the minimum of the original cost function \eqref{min cost} 
is given by 
\[
       J[u^*] = \lim_{T\rightarrow \infty} \frac{1}{T}\mean{J_0^*(\hat{x}_0)}
                         + \Tr(QV_\infty)
                  = \Tr[ P_\infty K_\infty D \Re(\Theta) D^\top K_\infty^\top] 
                         + \Tr(QV_\infty). 
\]
%


\section*{Appendix B: Condition for the cheap control \cite{Sivan,Seron Book,Seron}}

Let us consider the system whose transfer function matrix is given by 
$\Xi(s)=\bar{Q}(sI-A)^{-1}F$, which we simply call the system $(A, F, \bar{Q})$. 
First, if there exist a complex number $z\in{\mathbb C}$ and a vector 
$u$ such that $u^\top G(z)=0$ or $G(z)u=0$, then $z$ is called a {\it zero} 
(more precisely, it is called a transmission zero). 
Then the system $(A, F, \bar{Q})$ is called {\it minimum phase}, if all the 
zeros of $\Xi(s)$ have negative real part. 
Next the system $(A, F, \bar{Q})$ is called {\it right invertible}, if $\Xi(s)$ 
has full row rank for at least one $s\in{\mathbb C}$. 
In general, such a minimum phase and right invertible system is regarded 
as a system easy to control; 
an intuitive understanding of this fact is that there exists an ``inverse" and 
``stable" system (i.e., there exists $\Xi(s)^{-1}$ and all its poles have negative 
real part), and this system completely compensates $\Xi(s)$. 
In fact, as mentioned in Section 2.4, for such a system there exists a 
stabilizing controller that well suppresses the dynamical fluctuation of the 
(estimated) system variables.

Here we prove that the opto-mechanical system examined in Section~4 
actually satisfies the above condition for cheap control. 
Note that the LQG problem is now formulated with the choice 
\[
     \bar{Q}
        = \left[ \begin{array}{cccc}
             1 & 0 & 0 & 0 \\
             0 & 1 & 0 & 0 \\
           \end{array} \right], 
\]
which actually yields $Q=\bar{Q}^\top \bar{Q}={\rm diag}\{1, 1, 0, 0\}$; 
see Eq.~\eqref{min energy}. 
The $F$ matrix representing the actuator mechanism of the controller 
can be typically chosen as follows. 
First, if the oscillator can be directly manipulated via a piezo electrical 
device, then $F_1=[0, 1, 0, 0]^\top$, meaning that the momentum of the 
oscillator can be driven by an external force. 
Another typical setup for actuation is that the control is carried out 
by modulating the input probe field, in which case $F_2=[0,0,1,0]^\top$, 
where especially only the $\hat q_2$ quadrature is assumed to be modulated 
(it can be proven that modulating $\hat p_2$ does not affect on the 
condition to be fulfilled). 
Then we have 
\begin{eqnarray*}
& & \hspace*{-1em}
        \Xi_1(s)
           =\bar{Q}(sI-A)^{-1}F_1
           = \frac{1}{(s+\kappa/2)^2+\omega^2}
               \left[ \begin{array}{c}
                  \omega \\
                  s+\gamma/2 \\
               \end{array} \right],
\\ & & \hspace*{-1em}
        \Xi_2(s)
           =\bar{Q}(sI-A)^{-1}F_2
           = \frac{\lambda}{[(s+\kappa/2)^2+\omega^2](s+\kappa/2)}
               \left[ \begin{array}{c}
                  \omega \\
                  s+\gamma/2 \\
               \end{array} \right]. 
\nonumber
\end{eqnarray*}
Therefore, in both cases, the system $(A, F, \bar{Q})$ is minimum phase and 
right invertible.


\section*{Appendix C: Logarithmic negativity}

For a two-mode Gaussian state with mean zero, its correlation property can 
be completely characterized by the covariance matrix 
\begin{equation}
\label{V deconposition}
    V=\left[ \begin{array}{cc}
            V_1 & V_2 \\
            V_2^\top & V_3 \\
      \end{array} \right],
\end{equation}
where $V_i$ are $2\times 2$ matrices. 
In particular, the following {\it logarithmic negativity} \cite{Vidal,Plenio}
can be used as a reasonable measure of entanglement of this Gaussian state:
\begin{equation*}
    E_{\cal N}={\rm max}\big\{0,~-\log(2\nu)\big\},
\end{equation*}
where $\log x$ denotes the natural logarithm of $x$, and 
\[
    \nu=\frac{1}{\sqrt{2}}
         \sqrt{ \tilde{\Delta}
              -\sqrt{ \tilde{\Delta}^2-4{\rm det}(V)} },~~~
    \tilde{\Delta}={\rm det}(V_1)+{\rm det}(V_3)-2{\rm det}(V_2).
\]
Actually the state is entangled if and only if $E_{\cal N} > 0$.

In our case, the output of BS1 is an entangled Gaussian field; 
particularly when $\theta=\pi/2$, the covariance (more precisely the 
spectral density) matrix is given by Eq.~\eqref{V deconposition} with 
\begin{eqnarray*}
& & \hspace*{-1em}
     V_1 = {\rm diag}\{ \alpha_1^2 e^{r} + \beta_1^2, 
                       \alpha_1^2 e^{-r} + \beta_1^2 \} /2, 
\\ & & \hspace*{-1em}
     V_2 = {\rm diag}\{ \alpha_1\beta_1(1-e^{r}), \alpha_1\beta_1(1-e^{-r}) \} /2, 
\\ & & \hspace*{-1em}
     V_3 = {\rm diag}\{ \beta_1^2 e^{r} + \alpha_1^2, 
                        \beta_1^2 e^{-r} + \alpha_1^2 \} /2. 
\nonumber
\end{eqnarray*}
This yields $\nu=\sqrt{d-\sqrt{d^2-1}}/2$ with 
$d=2\alpha_1^2\beta_1^2(e^r+e^{-r}-2)+1$, and thus $E_{\cal N} > 0$ 
for all $\beta_1\in(0,1)$ and $r\neq 0$. 
Note that, hence, the maximal entangled field for $\theta=\pi/2$ is produced 
when $\alpha_1^2=\beta_1^2=1/2$.




\end{document}